# High-Resolution Spectroscopy of the Intermediate Impurity States near a Quantum Phase Transition

Yao Zhang[1], Tao Xie[1], Zhen-Yu Liu[1], Rui Wang[2,3], Wenhao Zhang[1], Chaofei Liu[1,*], and Ying-Shuang Fu[1,†]

[1]School of Physics and Wuhan National High Magnetic Field Center, Huazhong University of Science and Technology, Wuhan 430074, China

[2]National Laboratory of Solid State Microstructures and Department of Physics, Nanjing University, Nanjing 210093, China

[3]Collaborative Innovation Center for Advanced Microstructures, Nanjing University, Nanjing 210093, China

*cliu@hust.edu.cn
†yfu@hust.edu.cn

**ABSTRACT:** The intermediate behavior near a quantum phase transition is crucial for understanding the quantum criticality of various competing phases and their separate origins, yet remains unexplored for the multiple Yu-Shiba-Rusinov (YSR) states. Here, we investigated the detailed spectroscopic change of the exchange coupling-dependent YSR states near a quantum phase transition. The initially developed one pair of YSR states—induced by the Fe vacancy in monolayer Fe(Te,Se) superconductor, are clearly resolved with high resolution showing an evolution into two pairs of YSR peaks yet with dichotomy in their spectral features. Interestingly, while the lower-energy YSR branch enters the quantum phase transition region, the higher-lying one remains rigidly away from the lower-energy counterpart with a constant energy difference. Spectral weight analysis of the higher-energy branch yields an exponential dependence on the exchange coupling, which can be well rationalized by taking the two pairs of YSR states as a result of field splitting by the magnetic anisotropy. Our results unveil the intermediate region of a quantum phase transition with a magnetic anisotropy-induced splitting of the YSR resonance, and highlight a prospect for developing functional electronics based on the flexibly controllable multiple quantum states.

Impurity scattering has been widely studied in condensed matter at a fundamental level for discovering exotic physics phenomena. The example varies from Kondo screening [1] to Cooper pair breaking [2], which arise due to the scattering of itinerant electrons and Cooper pairs by impurities, respectively. Cooper pairs scattered by impurities will be induced with different types of in-gap states cooperatively decided by the pairing mechanism and the impurity properties [2]. Of particular interest is the magnetic impurities in *s*-wave superconductors; the induced in-gap bound states therein, usually called Yu-Shiba-Rusinov (YSR) states [3–5], have proven crucial in the atomic imaging of orbital states [6,7] and spin structure [8,9], together with the construction of Majorana fermions through engineered inter-impurity coupling [10–13]. In contrast, the Kondo effect will appear when the free electrons, instead of the Cooper pairs, are scattered by magnetic atoms in a metallic continuum [1]; the localized magnetic moment of the impurity atom is in turn screened by these normal-state electrons, and features a Kondo resonance at the Fermi level.

For the composite magnetic adsorbates/superconductors system, the competition between Cooper pairing and Kondo screening leads to two distinct ground states of a localized magnetic moment linked by a quantum phase transition (QPT) [14]. At the QPT regime, $J_c$ defines the critical Kondo exchange coupling between the magnetic adsorbate and superconductors. For weakly coupled case ($J<J_c$), the slightly perturbed superconducting condensation yields a reduced density of itinerant electrons available for Kondo screening. The preserved magnetic impurity will locally weaken the superconducting coherence accompanied with the in-gap YSR states, and meanwhile results in a free-spin ground state. By increasing the exchange coupling, the ground state changes from the free-spin to the Kondo-screened many-particle state (strongly coupled case: $J>J_c$). The two situations with different ground states can be distinguished by using the magnetic-anisotropy effect [15,16] or an applied magnetic field [16–20]. For a low-spin ($S_{imp} \leq 1$) impurity, the microscopic theory predicts divergent behaviors of YSR states in the presence of magnetic anisotropy. To be specific, for $J<J_c$ with a free-spin ground state, only single YSR state is expected despite the magnetic anisotropy-split multiplet ground states [16]. The reason lies in the exponentially decayed population, thus negligibly weak observable effect, for the higher-energy levels of the multiplet dictated by the Boltzmann distribution. For $J>J_c$ after QPT, the Kondo-screened singlet ground state naturally yields multiple YSR states. However, the intermediate behavior right near QPT has yet to be explored, partially suffering from the deficient ability in achieving a high resolution of the QPT region.

Here, we experimentally elucidated the QPT of YSR states induced by the Fe vacancies in monolayer Fe(Te,Se)/SrTiO$_3$ superconductor. Via gradually changing the tip-sample distance, a slowly varying gradient of the exchange coupling $J$ is accessible. The resulting subtle balance between Kondo screening and Cooper pairing allows a nearly continuous evolution of the ground state from the free-spin to the Kondo-screened situation. Using high-resolution scanning tunneling spectroscopy, the study carefully examined the changes of YSR states particularly in the vicinity of the QPT. Two pairs of closely entangled YSR states are observed evolving from the initially generated single pair. Remarkably, for such doubly-paired YSR states, while the lower-energy branch experiences the QPT, the higher-lying one appears throughout without a QPT. Furthermore, it remains robustly ~1 mV away from the corresponding lower-energy counterpart despite the considerable change of $J$. Assisted by detailed spectral-weight analysis, the doublet YSR states can be well explained by the field splitting of magnetic anisotropy. The results unravel the intermediate YSR states near a QPT, and demonstrate a delicate interplay between Kondo screening, Cooper pairing and magnetic anisotropy at the atomic scale.

The experiments were performed in a He3-refrigerator scanning tunneling microscope (STM) (Unisoku 1300) at 400 mK equipped with a molecular beam epitaxy (MBE) chamber (see Methods). Monolayer Fe(Te,Se)/SrTiO$_3$ is known as an interface-enhanced high-temperature superconductor. Figure 1(a) shows the surface topography of monolayer Fe(Te,Se) sample. The film spreading over the stepped SrTiO$_3$ is atomically uniform at sub-μm scale with high crystalline quality. The nominal stoichiometry of FeTe$_{1-x}$Se$_x$ with $x$=0.5 is inferred from the height of the second layer following previous studies [21]. Worm-like domain walls clearly seen on the surface [inset of Fig. 1(a)] are usually the preclude of the occurrence of superconductivity. Essentially, these domain walls arise as a relief of the stress generated due to the lattice mismatch between FeSe and SrTiO$_3$. Figure 1(b) displays the atomic-resolution view of the sample surface, revealing a square-ordered Se-terminated (001) lattice with irregularly distributed Se atom-like dimers. The dumbbell-shaped dimer defects with the center at Fe atom sites are reported to result from the Fe vacancies [6,22–24] carrying an effective magnetic moment [17]. At the reference defect-free region, the tunneling d$I$/d$V$ spectrum shows a fully depleted superconducting gap flanked by two pair of coherence peaks [red curve in Fig. 1(c)] typical of multi-band superconductivity. In contrast, at the Fe-vacancy sites, the in-gap YSR states are observed. Generally, the energy positions of the YSR states, as well as their numbers, vary among different sets of dimer defects [Fig. 1(c)], depending on the interplay among various factors, including the coupling strength $J$ with the underlying Fe(Te,Se) superconductor, the magnitude of their effective spin moments, and the total Kondo screening channels. The intensity asymmetry between the electron- ($V_s$>0) and hole-like ($V_s$<0) states is a reflection of the time-reversal symmetry breaking induced by the magnetic interactions between Fe vacancy and Fe(Te,Se).

We examined one of the Fe vacancies in detail [Fig. 2(a)]. The dumbbell-shaped Fe vacancy shows a two-fold symmetric local environment. The introduced scattering potential is expected to be anisotropic, and dictates a similarly anisotropic decay of coupling strength $J$ in space with respect to the defect site. STM provides a nanoscale spatial resolution, allowing the utilization of the tip's spatial sensitivity for detecting the local modulation of $J$. The Fe vacancy with purely one pair of YSR states is chosen to avoid the complication of multiple YSR peaks arising due to diverse origins. Figure 2(b) displays the 2D-conductivity plot of the line-cut spectra taken across such an Fe vacancy. A pair of in-gap YSR states are clearly visible near the center of Fe vacancy. In theory, when $J$ reaches a critical value, the two peaks of the energy-symmetric YSR states would move towards the middle position of the superconducting gap, forming a single peak at exactly zero energy [2,14,25,26]. Interestingly, under the influence of the defect-centered but spatially dispersive exchange coupling $J$, the paired YSR states gradually merge into a single state as moving away from the center of Fe vacancy [Fig. 2(b)]. The single peak is distinct from the topologically protected Majorana zero mode in the topological superconducting system [27]. Fundamentally, Majorana bound mode would be robustly pinned at zero energy despite the significant change of exchange coupling $J$. The $J$-varying result here represents a typical method in iron-based superconductors to distinguish the zero-energy states with topologically different nature.

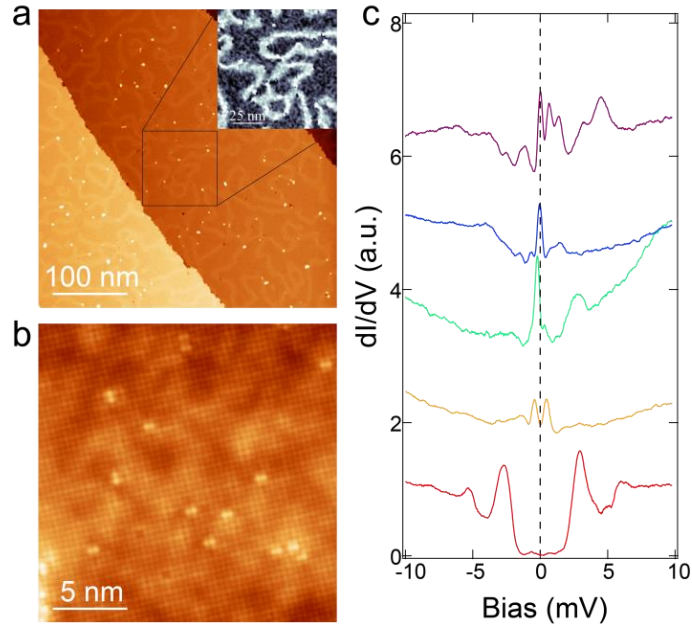

**Figure 1.** YSR states at the Fe vacancies in Fe(Te,Se). (a) STM morphology of monolayer Fe(Te,Se) film ($V_s$=1 V, $I_t$=10 pA) spanning multiple steps of SrTiO$_3$(001). (b) Atomically resolved STM image of monolayer Fe(Te,Se) ($V_s$=−100 meV, $I_t$=40 pA). (c) Representative tunneling spectra measured on different Fe vacancies. The superconducting gap spectrum (red curve) measured at the defect-free region is shown for comparison.

The intermediate YSR behavior near QPT is scrutinized via finely tunning the exchange coupling $J$. Notably, the doubly degenerate YSR states at Fermi energy [Fig. 2(b)] are essentially at a critical point of the QPT. To controllably tune $J$ for an in-depth investigation near the QPT, the tunneling barrier conductance $G_N$(=$I_t/V_s$), or the tip-sample distance, is quasi-continuously changed [28], essentially for modulating the electrostatic force between tip and defects [29–32]. Figure 2(c) shows the measured spectral evolution of a pair of YSR states for the Fe vacancy. The data are collected and ordered by successively approaching the tip towards the sample via increasing $G_N$ (from the top to the bottom spectrum). A collective shift of the YSR peaks can be seen crossing through the Fermi energy. In addition, as increasing $G_N$, the entire phase transition process appears undergoing a change from one to two pairs of YSR peaks. The convergence of the inner paired peaks at zero energy is indicative of a typical QPT phenomenon for the YSR states from free-spin to Kondo-screened ground state [33,34]. Across the quantum critical point, the increasingly larger exchange coupling $J$ dominates over the pairing energy. The resulting Cooper-pair breaking is characterized by intensity change of the lower-energy YSR states. Specifically, the larger-weight peak is rigidly shifted from the hole- to electron-like state [Fig. 2(c)]. The bias reversal of YSR intensity contrast represents the typical feature of a quantum transition of the ground state [35].

The intermediate impurity spectra are revealed with essentially two pairs of YSR states throughout the QPT process. Different from previous studies [15,37,38], the detailed YSR evolution near QPT regime is visualized here with a high resolution (~0.17 meV). Intriguingly, upon closer inspection, the splitting of YSR states is found emerging before (i.e., for $J<J_c$) the lower-energy YSR states undergoes the QPT [Fig. 2(c)]. While the YSR spectra near and after (i.e., for $J>J_c$) QPT show identifiable splitting, those far before QPT appear as only one pair of peaks. To uncover the emergence of the "new" pair of in-gap states quantitatively, we fitted the $G_N$-dependent YSR spectra using multi-Gaussian functions. The fitting procedure is capable of pinpointing the energy positions of all possible peaks involved in the YSR states. The best fitting results for the representative spectra before, near and after QPT are shown in Figs. 2(d)-2(f). Surprisingly, there are actually two pairs of YSR states (referred below as $p_1$-$p_4$ states) when the tip is furthest away from the sample. This is preliminarily hinted in their asymmetric lineshapes with respect to the center of each peak [Fig. 2(d)], where the intensity is selectively elevated at the high-absolute-energy side.

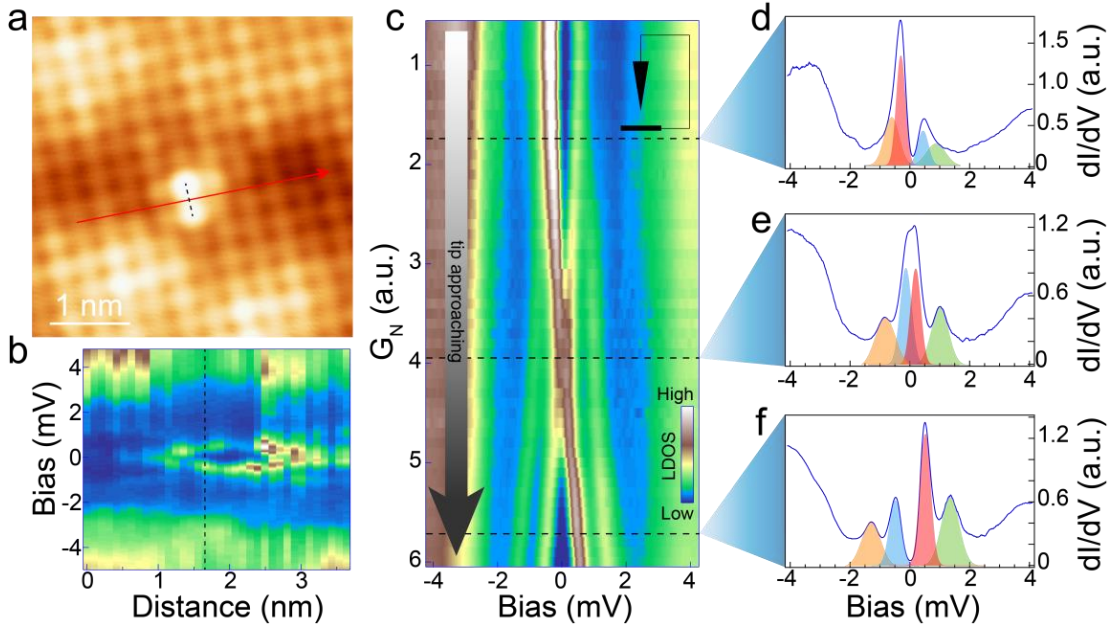

**Figure 2.** Detailed evolution of YSR states near the QPT. (a) Topographic image of the Fe vacancies (dimers) in Fe(Te,Se) ($V_s=-100$ meV, $I_t=100$ pA). (b) 2D display of the d$I$/d$V$ spectra taken along the arrow in (a). (c) 2D display of the normalized d$I$/d$V$ spectra with increasing tunneling conductance $G_N$ (from top to bottom) for the Fe vacancy in the center of (a). LDOS, local density of states. (d-f) Selected spectra at the dashed lines in (c), corresponding to three different cases: before, near and after the QPT. The shaded peaks highlight the components of the multi-Gaussian fit results for each spectrum near the Fermi level.

Our results are found closely near the QPT to an extent where the deep free-spin ground-state regime—with the Kondo-screened level being well above the doublet ground state [e.g., Fig. 3(b)]—is absent. By summarizing the multi-Gaussian fitting results for $G_N$-dependent spectra, Fig. 3(a) sketches the evolution of the energy positions of YSR states as increasing the exchange coupling $J$. In general, the behavior of YSR states tuned by $J$ is shaped by the combined influences of the degeneracy of ground or excited state, the magnetic anisotropy, and the selection rules, as detailed in the following:

(i) Firstly, the low-symmetry (two-fold) local environment of the Fe vacancy is expected to generate magnetocrystalline anisotropy. Theoretically, the magnetic anisotropy is effectively considered by the spin Hamiltonian $H_{ani.} = DS_z^2 + E(S_x^2 - S_y^2)$ in the Anderson impurity model [36]. Here, $D$ and $E$ terms describe the axial and transverse anisotropies. For sufficiently small $J$, the Kondo energy scale is smaller than the Cooper-pairing energy. Under the influence of magnetic anisotropy, the resulting ground state is calculated to be the unscreened spin-split doublet ($S_z=0, \pm1$, for $S_{imp}=1$), while the excited state is a Kondo-screened singlet ($S_{imp}'=1/2$) [Fig. 3(b)] [19]. Here we have taken the impurity spin $S_{imp}=1$ for example and considered only the $D$ term.

(ii) Secondly, according to the microscopic theory, the YSR states are induced by the excitation and radiation processes between the ground and the excited states with a selection rule of altered spin $\Delta S_z=\pm1/2$. Thus, the YSR states originating from the electron transition between $S_z=1/2$ to 0 and 1 are allowed, while the one between 0 and $\pm1$ is forbidden.

(iii) Lastly, by further considering the Boltzmann distribution at 400 mK, the thermal occupation of $S_z=0$ state is exponentially decayed. Such a thermal effect results in an ultra-low probability of the excitation between $S_z=1/2$ and 0.

Taken together, under sufficiently small $J$ [Fig. 3(b)], the higher-energy YSR states corresponding to $S_z=1\to1/2$ excitation would be probed with a pair of lower-intensity satellite YSR peaks—that is induced by $S_z=0\to1/2$ excitation and positioned at the lower-energy side. This is obviously different from the experimental measurements showing instead higher-energy satellite peaks [green and orange shaded peaks in Fig. 2(d)]. The absence of the region with deep free-spin ground state [i.e., Fig. 3(b)] is also evidenced in the considerably large $J$ in our experiments, as reflected by the far distance of the observed YSR states to the superconducting gap edges.

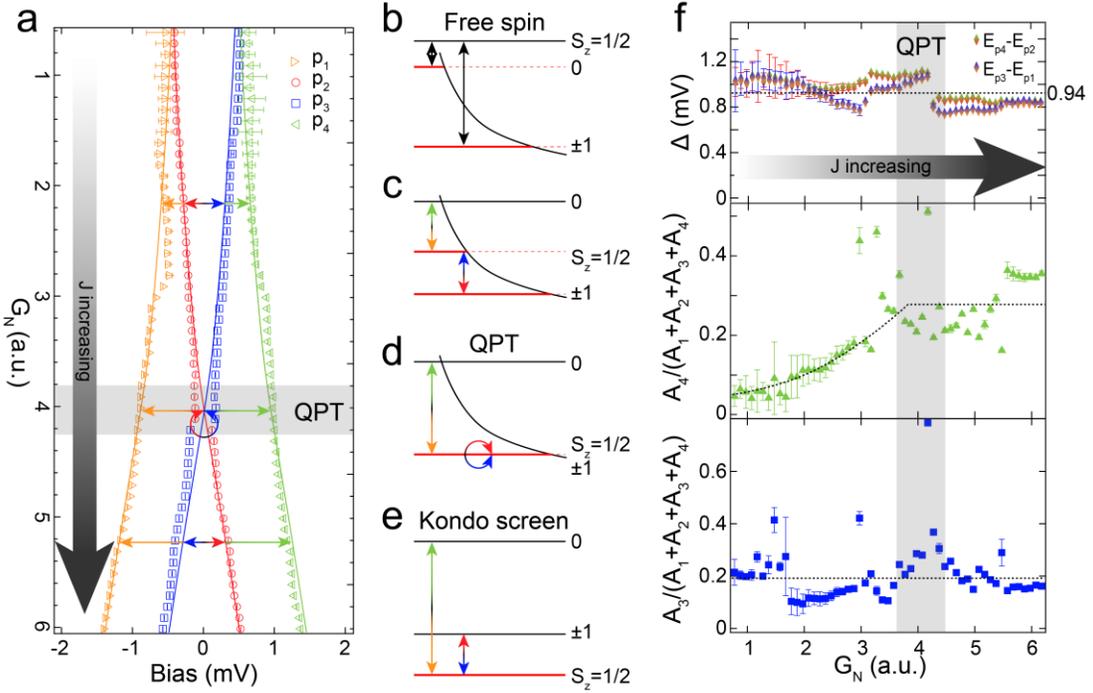

**Figure 3.** Physics scenario of the QPT. (a) Summarized energy positions of the YSR states (numbered as $p_1$-$p_4$) for the spectra in Fig. 2(c) extracted via multi-Gaussian fits. Thin solid lines are shown as guidelines. (b-e) Schematics of the exchange coupling-tunable transition of the ground state from free-spin to Kondo-screened case for $S_{imp}=1$. The influence of magnetic anisotropy and Boltzmann distribution at finite temperature (curved lines) have been considered. (f) Evolution of the spacing $\Delta$ of YSR states ($\Delta = E_{p_4} - E_{p_2} = E_{p_3} - E_{p_1}$) and the scaled peak area $A_{3,4}/A_{tot}$ ($A_{tot}=A_1+A_2+A_3+A_4$) for $p_3$ and $p_4$ states with increasing tunnel coupling $G_N$.

The YSR behavior near QPT is well described within a scenario incorporating the magnetic anisotropy. Throughout the following discussion, the introduced energy-leveling formulism influenced by the magnetic anisotropy [Figs. 3(c)-3(e)] is adopted. As mentioned above, the starting point in the $J$ range accessible in our data yields two pairs of YSR states with nearly degenerate energies [e.g., Fig. 2(d)]. This corresponds to a regime where the $S_z=1/2$ level is located near the middle position between $S_z=0$ and $\pm 1$ states [a crossover regime; Fig. 3(c)]. The close proximity of $S_z=1/2$ and $\pm 1$ indicates that the effective spin of the Fe vacancy is already coupled to the substrate with considerable strength, even without the tip-regulated coupling. As the coupling strength $J$ increases with the tip approaching, the $S_z=1/2$ state gradually decreases towards the $S_z=\pm 1$ ground state. The selection rule and Boltzmann distribution combined now give rise to an energy of the low-probability excitation process ($S_z=1/2\to 0$) exceeding the energy of the main excitation ($S_z=1\to 1/2$) [c.f. Fig. 3(c)]. This produces YSR spectra with satellite peaks on the higher-energy side of the main peaks consistently as observed [Fig. 2(d)], supporting the proposed interpretation by considering the magnetic anisotropy.

We now discuss the processes near and after ($J\gtrsim J_c$) the QPT following the magnetic-anisotropy framework. When the $S_z=1/2$ state is reduced to the energy position of $S_z=\pm 1$, the system reaches a special case where the QPT emerges [Fig. 3(d)]. At the QPT point, the excitation between $S_z=1$ and $1/2$ decreases to zero energy and yields a single YSR resonance state at the Fermi level. Meanwhile, the excitation energy between $S_z=1/2$ and $0$ is exactly equal to the level spacing of the magnetic anisotropy-split doublet ($S_z=0, \pm 1$). In the strongly coupled Kondo-screened state after QPT, the two excitation processes from the ground state $S_z=1/2$ to the excited states $S_z=0, 1$ [Fig. 3(e)] would result in two pairs of YSR states with an expected $J$-independent energy difference of $D$ scale. Quantitatively, the YSR energy difference $\Delta$ corresponding to the energy scale of magnetic anisotropy $D$ was extracted from the multi-Gaussian fitting results [top panel of Fig. 3(f)]. $\Delta$ was found to indeed remain constant after the QPT, and more universally, robustly unchanged throughout the entire QPT. The constant energy difference ($\Delta \sim 0.94$ meV) between the two sets of paired YSR branches provides a measure of the magnetic-anisotropy energy $D$. This in turn highlights the two pairs of YSR states with a "locked" nature as predicted for the magnetic-anisotropy picture. Alternative explanations beyond magnetic anisotropy are thus unlikely if involving multiple independent Kondo-screening channels as discussed later.

The double YSR states with an origin of magnetic anisotropy is quantitatively corroborated by the spectral weight analysis. In principle, the spectral weight of a YSR state, or its peak area, is determined by the excitation probability that is proportional to the electron population [15]. As the $S_z=1/2$ level decreases with increasing $J$, its electron occupation increases exponentially following the Boltzmann distribution law, while the ground state $S_z=1$ always remains fully occupied. Here, we consider only the regime after $S_z=1/2$ level crosses the middle position of $S_z=0, \pm 1$ states [e.g., Figs. 3(c) and 3(d)]. Correspondingly, the YSR state for $S_z=1/2 \rightarrow 0$ transition ($p_4$ state) is expected to show an exponential increase in spectral weight. In contrast, the spectral weight of the YSR state for $S_z=1 \rightarrow 1/2$ transition ($p_3$ state) will be constant. After the QPT, the $S_z=1/2$ level becomes the ground state with full occupation. The YSR states for the same $S_z=1/2 \rightarrow 0$ transition ($p_4$ state) would hence saturate without further increase of the spectral weight. The other YSR state for $S_z=1/2 \rightarrow 1$ ($p_3$ state)—still a transition between the ground and the excited states, show no change as compared with that before the QPT regarding spectral weight. This picture is justified in Fig. 3(f) (middle and bottom panels), where we plot the $J$ dependence of $A_3$ and $A_4$. In detail, $A_3$ ($A_4$) denotes the rescaled spectral weight for the $p_3$ ($p_4$) YSR state corresponding to $S_z=1 \leftrightarrow 1/2$ ($1/2 \rightarrow 0$) transition. Strikingly, as increasing $J$, $A_4$ increases exponentially, and after the QPT, nearly saturates with a fluctuating background. Nevertheless, $A_3$ presents a trend with independence of $J$ across the QPT. Therefore, the result exclusively reveals the role of magnetic anisotropy played in splitting the free-spin doublet to induce the satellite YSR states ($p_1$ and $p_4$). By extending the spectral evolution towards the smaller-$J$ region near and between those sketched in Figs. 3(b) and 3(c), another QPT is expectable, which serves as a prediction of the magnetic-anisotropy interpretation left for future check in experiments.

The YSR occupation of a Boltzmann-distributed nature excludes other explanations that incorporate multiple independent Kondo-screening channels. While the spectral signals of multiple YSR states are similar, their formation mechanisms can be diverse. In principle, double scattering channels with different angular momenta [37,38] or $d$ orbitals can also give rise to two pairs of YSR states [16]. Yet, it is crucial to emphasize that these two channels are independent, for which a synchronous shift of the two pairs of YSR states as observed is unfavored. Within the multi-channel scenario, to induce only two pairs of YSR states, the free-spin doublet in each channel has to be degenerate. Thus, both the free-spin doublet and the Kondo-screened singlet would be fully occupied or empty when serving as the state in turn for inducing YSR excitations. The resulting YSR states would show no Boltzmann distribution-associated exponential dependence on $J$, in an apparent contradiction with our observation.

In summary, we investigated the intermediate impurity states near the QPT regime with high spectral resolution. Two pairs of entangled YSR states with a suggested magnetic-anisotropy origin are revealed with dichotomy in their behaviors across the QPT. Further studies can be extended to the Fe vacancy dimer, in which the atom-like YSR bound states of the two Fe vacancies will hybridize and split into bonding and anti-bonding states, provided that their magnetic moments are parallel. This represents a fascinating direction to study how the competition between exchange coupling and orbital hybridization will affect the behavior of YSR states. In addition, as demonstrated in our work, the Kondo-screened singlet was tuned within a small range near the lower branch of the magnetic anisotropy-split doublet [Figs. 3(c)-3(e)]. Such an exchange coupling-tunable leveling of the singlet—turning out being below or above the doublet, provides a controllable method to flexibly change the ground states between Kondo-screened singlet and free-spin doublet. The easy switch of ground states shows the prospect for constructing qubits based on the singlet-doublet states. By further establishing an ordered YSR impurity lattice, a new platform may be developed for the new-generation quantum computation.

■ ACKNOWLEDGMENTS

This work is funded by the National Key Research and Development Program of China (Grants No. 2022YFA1402400, and 2018YFA0307000), the National Science Foundation of China (Grants No. 92265201, U20A6002, 12174131, 12088101, 12047508, and 11974422), the Natural Science Foundation of Hubei (Grant No. 2022CFB033), and the Knowledge Innovation Program of Wuhan-Basic Research (Grant No. 2023010201010056).

■ REFERENCES

[1] J. Kondo, *Resistance Minimum in Dilute Magnetic Alloys*, Prog. Theor. Phys. **32**, 37 (1964).

[2] A. V. Balatsky, I. Vekhter, and J.-X. Zhu, *Impurity-Induced States in Conventional and Unconventional Superconductors*, Rev. Mod. Phys. **78**, 373 (2006).

[3] L. Yu, *Bound state in superconductors with paramagnetic impurities*, Acta Phys. Sin. **21**, 75 (1965).

[4] H. Shiba, *Classical Spins in Superconductors*, Prog. Theor. Phys. **40**, 435 (1968).

[5] A. I. Rusinov, *On the Theory of Gapless Superconductivity in Alloys Containing Paramagnetic Impurities*, Zh. Eksp. Teor. Fiz. **56**, 2047 (1969).

[6] H. Zhang, Z. Ge, M. Weinert, and L. Li, *Sign Changing Pairing in Single Layer FeSe/SrTiO$_3$ Revealed by Nonmagnetic Impurity Bound States*, Commun. Phys. **3**, 75 (2020).

[7] Q. Fan et al., *Plain s-Wave Superconductivity in Single-Layer FeSe on SrTiO$_3$ Probed by Scanning Tunnelling Microscopy*, Nat. Phys. **11**, 946 (2015).

[8] L. Cornils, A. Kamlapure, L. Zhou, S. Pradhan, A. A. Khajetoorians, J. Fransson, J. Wiebe, and R. Wiesendanger, *Spin-Resolved Spectroscopy of the Yu-Shiba-Rusinov States of Individual Atoms*, Phys. Rev. Lett. **119**, 197002 (2017).

[9] D. Wang, J. Wiebe, R. Zhong, G. Gu, and R. Wiesendanger, *Spin-Polarized Yu-Shiba-Rusinov States in an Iron-Based Superconductor*, Phys. Rev. Lett. **126**, 076802 (2021).

[10] T.-P. Choy, J. M. Edge, A. R. Akhmerov, and C. W. J. Beenakker, *Majorana Fermions Emerging from Magnetic Nanoparticles on a Superconductor without Spin-Orbit Coupling*, Phys. Rev. B **84**, 195442 (2011).

[11] F. Pientka, L. I. Glazman, and F. von Oppen, *Topological Superconducting Phase in Helical Shiba Chains*, Phys. Rev. B **88**, 155420 (2013).

[12] S. Nakosai, Y. Tanaka, and N. Nagaosa, *Two-Dimensional p-Wave Superconducting States with Magnetic Moments on a Conventional s-Wave Superconductor*, Phys. Rev. B **88**, 180503 (2013).

[13] S. Nadj-Perge, I. K. Drozdov, J. Li, H. Chen, S. Jeon, J. Seo, A. H. MacDonald, B. A. Bernevig, and A. Yazdani, *Observation of Majorana Fermions in Ferromagnetic Atomic Chains on a Superconductor*, Science **346**, 602 (2014).

[14] K. J. Franke, G. Schulze, and J. I. Pascual, *Competition of Superconducting Phenomena and Kondo Screening at the Nanoscale*, Science **332**, 940 (2011).

[15] N. Hatter, B. W. Heinrich, M. Ruby, J. I. Pascual, and K. J. Franke, *Magnetic Anisotropy in Shiba Bound States across a Quantum Phase Transition*, Nat. Commun. **6**, 8988 (2015).

[16] R. Žitko, O. Bodensiek, and T. Pruschke, *Effects of Magnetic Anisotropy on the Subgap Excitations Induced by Quantum Impurities in a Superconducting Host*, Phys. Rev. B **83**, 054512 (2011).

[17] W. Li et al., *Phase Separation and Magnetic Order in K-Doped Iron Selenide Superconductor*, Nat. Phys. **8**, 126 (2012).

[18] T. Machida, Y. Nagai, and T. Hanaguri, *Zeeman Effects on Yu-Shiba-Rusinov States*, Phys. Rev. Res. **4**, 033182 (2022).

[19] T. Zhang et al., *Phase Shift and Magnetic Anisotropy Induced Field Splitting of Impurity States in (Li$_{1-x}$Fe$_x$)OHFeSe Superconductor*, Phys. Rev. Lett. **130**, 206001 (2023).

[20] S. Karan, H. Huang, A. Ivanovic, C. Padurariu, B. Kubala, K. Kern, J. Ankerhold, and C. R. Ast, *Tracking a Spin-Polarized Superconducting Bound State across a Quantum Phase Transition*, Nat. Commun. **15**, 459 (2024).

[21] Y. Liu et al., *Spatial Inhomogeneity of Superconducting Gap in Epitaxial Monolayer FeTe$_{1-x}$Se$_x$ Films*, Phys. Rev. B **108**, 214514 (2023).

[22] D. Huang, T. A. Webb, C.-L. Song, C.-Z. Chang, J. S. Moodera, E. Kaxiras, and J. E. Hoffman, *Dumbbell Defects in FeSe*

*Films: A Scanning Tunneling Microscopy and First-Principles Investigation*, Nano Lett. **16**, 4224 (2016).

[23] G. Gong *et al.*, *Oxygen Vacancy Modulated Superconductivity in Monolayer FeSe on SrTiO$_3$*, Phys. Rev. B **100**, 224504 (2019).

[24] K. Bu, W. Zhang, Y. Fei, Y. Zheng, F. Ai, Z. Wu, Q. Wang, H. Wo, J. Zhao, and Y. Yin, *Observation of an Electronic Order along [110] Direction in FeSe*, Nat. Commun. **12**, 1385 (2021).

[25] R. S. Deacon, Y. Tanaka, A. Oiwa, R. Sakano, K. Yoshida, K. Shibata, K. Hirakawa, and S. Tarucha, *Tunneling Spectroscopy of Andreev Energy Levels in a Quantum Dot Coupled to a Superconductor*, Phys. Rev. Lett. **104**, 076805 (2010).

[26] J. Bauer, J. I. Pascual, and K. J. Franke, *Microscopic Resolution of the Interplay of Kondo Screening and Superconducting Pairing: Mn-Phthalocyanine Molecules Adsorbed on Superconducting Pb(111)*, Phys. Rev. B **87**, 075125 (2013).

[27] S. Nadj-Perge, I. K. Drozdov, B. A. Bernevig, and A. Yazdani, *Proposal for Realizing Majorana Fermions in Chains of Magnetic Atoms on a Superconductor*, Phys. Rev. B **88**, 020407 (2013).

[28] P. Fan *et al.*, *Observation of Magnetic Adatom-Induced Majorana Vortex and Its Hybridization with Field-Induced Majorana Vortex in an Iron-Based Superconductor*, Nat. Commun. **12**, 1348 (2021).

[29] M. Ternes, C. P. Lutz, C. F. Hirjibehedin, F. J. Giessibl, and A. J. Heinrich, *The Force Needed to Move an Atom on a Surface*, Science **319**, 1066 (2008).

[30] M. Ternes, C. González, C. P. Lutz, P. Hapala, F. J. Giessibl, P. Jelínek, and A. J. Heinrich, *Interplay of Conductance, Force, and Structural Change in Metallic Point Contacts*, Phys. Rev. Lett. **106**, 016802 (2011).

[31] L. Farinacci, G. Ahmadi, G. Reecht, M. Ruby, N. Bogdanoff, O. Peters, B. W. Heinrich, F. Von Oppen, and K. J. Franke, *Tuning the Coupling of an Individual Magnetic Impurity to a Superconductor: Quantum Phase Transition and Transport*, Phys. Rev. Lett. **121**, 196803 (2018).

[32] L. Malavolti *et al.*, *Tunable Spin–Superconductor Coupling of Spin 1/2 Vanadyl Phthalocyanine Molecules*, Nano Lett. **18**, 7955 (2018).

[33] K. Satori, H. Shiba, O. Sakai, and Y. Shimizu, *Numerical Renormalization Group Study of Magnetic Impurities in Superconductors*, J. Phys. Soc. Jpn. **61**, 3239 (1992).

[34] O. Sakai, Y. Shimizu, H. Shiba, and K. Satori, *Numerical Renormalization Group Study of Magnetic Impurities in Superconductors. II. Dynamical Excitation Spectra and Spatial Variation of the Order Parameter*, J. Phys. Soc. Jpn. **62**, 3181 (1993).

[35] M. Ruby, F. Pientka, Y. Peng, F. von Oppen, B. W. Heinrich, and K. J. Franke, *Tunneling Processes into Localized Subgap States in Superconductors*, Phys. Rev. Lett. **115**, 087001 (2015).

[36] N. Tsukahara *et al.*, *Adsorption-Induced Switching of Magnetic Anisotropy in a Single Iron(II) Phthalocyanine Molecule on an Oxidized Cu(110) Surface*, Phys. Rev. Lett. **102**, 167203 (2009).

[37] C. P. Moca, E. Demler, B. Jankó, and G. Zaránd, *Spin-Resolved Spectra of Shiba Multiplets from Mn Impurities in MgB$_2$*, Phys. Rev. B **77**, 174516 (2008).

[38] S.-H. Ji, T. Zhang, Y.-S. Fu, X. Chen, X.-C. Ma, J. Li, W.-H. Duan, J.-F. Jia, and Q.-K. Xue, *High-Resolution Scanning Tunneling Spectroscopy of Magnetic Impurity Induced Bound States in the Superconducting Gap of Pb Thin Films*, Phys. Rev. Lett. **100**, 226801 (2008).